\documentclass[fleqn,10pt]{wlscirep}
\usepackage{graphicx,color}

\newcommand{\mean}[1]{\langle #1 \rangle}

\newcommand{\phim}{\phi_\text{m}}
\newcommand{\phif}{\phi_\text{f}}
\newcommand{\phig}{\phi_\text{g}}
\newcommand{\tbr}{\tau_\text{B}}
\newcommand{\tx}{\tau_\text{x}}
\newcommand{\tf}{\tau_\text{f}}
\newcommand{\sr}{\dot\gamma}
\newcommand{\nc}{n_\text{c}}
\newcommand{\ns}{n_\text{s}}
\newcommand{\pb}{P_\text{B}}

\begin{document}

\title{The role of shear in crystallization kinetics: \\ From suppression to enhancement}

\author{David Richard}
\author{Thomas Speck}
\affil{Institut f\"ur Physik, Johannes Gutenberg-Universit\"at Mainz, Staudingerweg 7-9, 55128 Mainz, Germany}

\begin{abstract}
  In many technical applications, but also in natural processes like ice nucleation in clouds, crystallization proceeds in the presence of stresses and flows, hence the importance to understand the crystallization mechanism in simple situations. We employ molecular dynamics simulations to study the crystallization kinetics of a nearly hard sphere liquid that is weakly sheared. We demonstrate that shear flow both enhances and suppresses the crystallization kinetics of hard spheres. The effect of shear depends on the quiescent mechanism: suppression in the activated regime and enhancement in the diffusion-limited regime for small strain rates. At higher strain rates crystallization again becomes an activated process even at densities close to the glass transition.
\end{abstract}

\flushbottom
\maketitle
\thispagestyle{empty}


\section*{Introduction}

The making of a material is a process that often involves phase transformations. How a material transforms from one stable phase to another has wide-spread consequences: for technological applications such as metal alloys~\cite{porter}, colloids~\cite{ande02,tan14,peng15}, and the graphite-diamond transition of carbon~\cite{khal11}; but also for the nucleation of ice in clouds~\cite{virt10} affecting our climate. Phase transformations have been studied primarily after a sudden change of thermodynamic conditions but with dynamics that obey detailed balance. More realistically however, such transformations proceed away from thermal equilibrium and in the presence of external mechanical stresses and flows, for example when preparing colloidal crystals through spin-coating~\cite{jian04}. To study the basic physical mechanisms, here we consider as a model system a supersaturated liquid of hard spheres~\cite{auer01,Tanaka10,dijkstra11} and generate a non-vanishing stress through constantly shearing the liquid.


Hard spheres (or more generally, hard particles~\cite{dama12}) are a paradigm for a wide range of soft materials in which entropy dominates over energetic forces and is one of the best-studied model systems in statistical mechanics. Moreover, many aspects of liquids are dominated by the packing of atoms. In the absence of shear, even though there are no direct forces between particles, hard spheres crystallize at sufficiently high packing fractions $\phi>\phif$ with $\phif\simeq0.492$ when the entropy of the crystal has become higher than that of the disordered liquid. Experimentally, the limit of true hard spheres can be approached using colloidal suspensions~\cite{Pal14}, which allow the direct observation of nucleation events~\cite{gass01,tan14}. The exact microscopic pathway from liquid to solid remains the focus of intensive research. The classical picture has been that of a one-step processes in which a rare fluctuation leads to densification and the emergence of a nucleus with the same symmetry as the solid~\cite{beck35}. More recent work has observed different pre-cursors during the initial stage that have a different symmetry than the final solid. This leads to a picture of crystallization as a two-step~\cite{Tanaka10,Snook10}, or even three-step~\cite{tan14}, process. Increasing the density, the entropic barrier to nucleation shrinks with dynamics that effectively become arrested on experimental time scales for packing fractions $\phi>\phig$ with $\phig\simeq0.58$. How the crystallization pathway then changes is currently under investigation~\cite{sanz14}.

Weakly sheared liquids still crystallize but with kinetics that are significantly altered. Understanding and controlling these kinetics at least qualitatively is of technological importance in the search of new protocols to process materials. For example, growing crystals without defects requires a small growth rate once the critical nucleus has formed. Typically this implies conditions close to coexistence~\cite{ande02}, where, however, the nucleation time (the time for the critical nucleus to appear) becomes prohibitively large. No consensus has emerged so far regarding the role of shear in numerical~\cite{Blaak2004,Barrat08,cerd08,land13} and experimental studies~\cite{Dhont05,Imhof09,Zhen15} with and without attractive interactions. On one hand the suppression of crystallization has been reported for weak, moderate, and high strain rates~\cite{Blaak2004,Imhof09,cerd08}, which is characterized by a decrease of the structural order in the sheared crystal~\cite{Imhof09}, larger critical nuclei and nucleation barriers compared to the quiescent case~\cite{Blaak2004}, and finally the break-up and melting of solid clusters~\cite{cerd08}. On the other hand, weak shear can induce crystallization in the disordered liquid and glassy state~\cite{Barrat08,Zhen15}. To complicate matters, an optimal strain rate has been reported~\cite{Dhont05,cerd08,land13}, where at first crystallization is promoted but hampered at larger strain rates. We will show that all of these features are reproduced in a sheared hard sphere liquid.


\begin{figure*}
  \centering
  \includegraphics[scale=1.0]{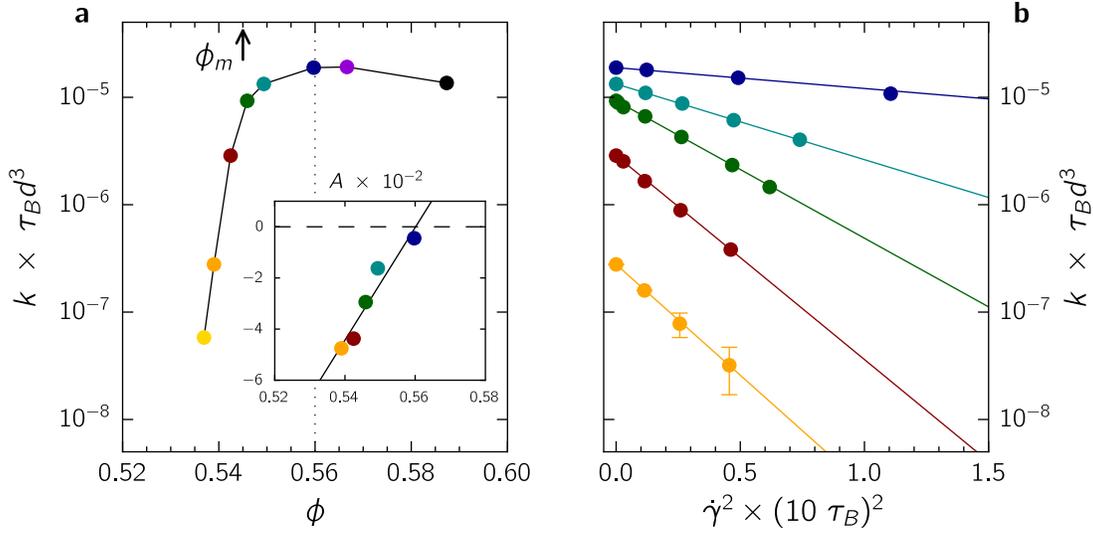}
  \caption{Crystallization kinetics. (a)~Crystallization rate $k$ \emph{vs.} the (effective) packing fraction $\phi$ for $\sr=0$. The arrow indicates the melting point $\phim\simeq0.545$ of hard spheres. (b)~Crystallization rate $k$ \emph{vs.} the square of the strain rate $\sr^2$ for the different packing fractions (colors agree with a). The solid lines are fits to Equation~(\ref{eq:rate}). Inset: The fitted values for $A$ as a function of packing fraction $\phi$, where the solid line is $A\propto(\phi-\phi_0)$ with $\phi_0\simeq 0.56$. If not shown then error bars in this figure are smaller than the symbol size.\label{fig:rate}}
\end{figure*}

\section*{Results}

As has been done in previous studies~\cite{Tanaka10,dijkstra11}, we do not simulate true hard spheres but particles interacting pairwise via the Weeks-Chandler-Andersen potential, which is mapped onto hard spheres through an effective diameter $d$ (see \emph{Methods} for details and definitions). We calculate the crystallization rate density $k$ (in Brownian units $\tbr d^3$) for different values of the packing fraction $\phi$ and strain rate $\sr$. To this end we bracket the crystallization time $\tx=(\tx^<+\tx^>)/2$ (the spread is included in the error estimation) by the two times $\tx^<$ ($\tx^>$) such that $n<\ns$ ($n>\ns$) for all $t<\tx^<$ ($t>\tx^>$). The exact value $\nc<\ns\ll N$ of the threshold is not important as long as it is larger than the critical nucleus size $\nc$ and much smaller than the system size. We found that $\ns=200$ is a convenient value, which in the following is used for all packing fractions studied. For the lowest density we have performed a committor analysis~\cite{Bol02,Dell11} and checked that the probability to commit to the solid state is unity for $n\geqslant200$. For the highest density $n\simeq200$ roughly corresponds to the onset of the linear growth regime. From 100 independent runs, the crystallization rate is then determined as $k=(V\mean{\tau_x})^{-1}$.

In the quiescent case ($\sr=0$) the rate is a non-monotonic function of the packing fraction, see Fig.~\ref{fig:rate}a: It has a maximum around $\phi\simeq0.56$ and is decreasing fast when going to lower densities (lower supersaturation). The calculated rates agree well with experimental results~\cite{dijkstra11,Pal14}. Note that densities even closer to freezing can be studied using rare-event sampling methods~\cite{auer01,dijkstra11} but predict absolute rates that are several orders of magnitude too small compared to experimental results~\cite{Pal14}. A comprehensive explanation of this discrepancy is still an open issue.

The crystallization rate also decreases at higher densities. This non-monotonicity reflects the turnover from an activated process at lower densities, where a sufficiently large solid cluster has to be nucleated to overcome interfacial tension, to a crystallization process that is limited by diffusion~\cite{Megen95}. In colloidal suspensions and complex fluids it is the mobility of particles that is rate-limiting, while in atomistic fluids it is the conduction of the released latent heat.

\subsection*{Activated regime}

We now apply a steady shear flow leading to a non-vanishing stress. The shear flow is weak so that for the range of strain rates studied the system remains in the linear response regime and the temperature is approximately constant. Since we consider a non-equilibrium situation we require a thermostat to remove the dissipated heat. We have been particularly careful in selecting this thermostat in order to minimize the deviation of the kinetic energy from the bath temperature while not perturbing collective dynamics. Note that the strain rates studied here are much smaller than what is required for shear-induced ordering~\cite{verm05}.

In Fig.~\ref{fig:rate}b, we observe for packing fractions $\phi\leqslant0.56$ a strong suppression of crystallization. The rate decreases as a function of the square of the strain rate,
\begin{equation}
  \label{eq:rate}
  \ln k(\phi,\sr) \approx \ln k_0(\phi) + A(\phi)\sr^2,
\end{equation}
where $k_0$ is the quiescent rate. Such a quadratic dependence to lowest order is expected from symmetry considerations alone since the rate should be invariant under the inversion ($\sr\mapsto-\sr$) of the flow profile. Moreover, the fitted expansion coefficients $A<0$ for the lowest densities are well described by a linear function $A\propto(\phi-\phi_0)$ with $\phi_0\simeq0.56$ well below the glassy regime. Given the strong dependence of the quiescent rate $k_0$ on density, this simple result is somewhat surprising. The density $\phi_0$ approximately indicates the crossover from a regime where crystallization is suppressed by shear flow to a regime where it is enhanced with $A>0$ (at least for weak shear). The location of this crossover agrees with the location for the qualitative change from activated to diffusion-limited.


\subsection*{Committor analysis}

A good reaction coordinate to unambiguously describe the process of crystallization is the probability $\pb$ to commit to the solid state. To this end we perform a committor analysis~\cite{Bol02,Dell11}. $\pb$ is computed from several ($20$ in our case) fleeting trajectories -- short runs of length $\tf$ -- for every configuration along a stored trajectory with randomized velocities. The fleeting time $\tf\simeq 18\tbr\ll\tx$ is much shorter than the nucleation time $\tx$ with mean $\mean{\tx}\simeq 750\tbr$, but large enough to allow the system to unambiguously reach the threshold ($\ns=200$). The committor for one configuration is calculated as the ratio between the number of fleeting trajectories that reach the solid divided by the total number of trials fired from that configuration. Such an analysis can also be performed in a driven system. Some care has to be taken regarding the initial velocities, which here are taken from the equilibrium Maxwell-Boltzmann distribution. Due to the small strain rates, the perturbation of the steady state is negligible compared to the fleeting time $\tf$.

\begin{figure}
  \centering
  \includegraphics[scale=1.0]{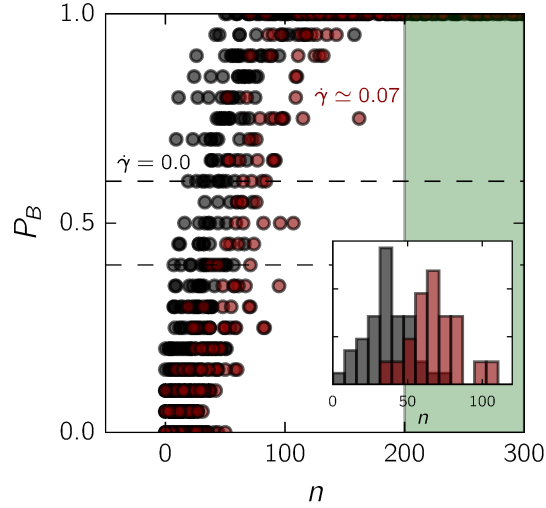}
  \caption{Committor analysis for packing fraction $\phi\simeq0.539$. Scatter plot showing the correlations between the committor $\pb$ and the size $n$ of the largest cluster for the quiescent (gray) and shear driven case (red) with strain rate $\sr\tbr\simeq0.07$. The shaded area indicates $n\geqslant\ns=200$, for which the probability that solid clusters continue to grow has reached $\pb\simeq1$. The dashed lines indicate the configurations that constitute the transition state ensemble, for which the distribution of nucleus sizes is shown in the inset.\label{fig:com}}
\end{figure}

Fig.~\ref{fig:com} shows for the lowest density $\phi\simeq0.539$ studied that the correlations between $\pb$ and the cluster size $n$ are similar for both the quiescent and the sheared liquid. For the latter, equal $\pb$ typically have larger cluster sizes which indicates that smaller clusters are more likely to break up under shear. This intuitively agrees with the observation of a suppression of nucleation. Of particular interest are configurations for which $\pb\sim0.5$, \emph{i.e.}, for which the chances to fall back to the liquid or to surmount the barrier are equal. These configurations constitute the transition state ensemble, for which the distribution of cluster sizes are shown in the inset of Fig.~\ref{fig:com}. The mean values $\nc\simeq36$ for $\sr=0$ and $\nc\simeq68$ for $\sr\tbr\simeq0.07$ are estimates of the critical cluster sizes for the quiescent and sheared liquid, respectively.

\subsection*{Effective chemical potential}

Classical nucleation theory predicts a crystallization rate $k_0=\kappa\exp(-\Delta G/k_\text{B}T)$ with two contributions: a kinetic pre-factor $\kappa$ and an exponential free energy barrier $\Delta G\propto|\Delta\mu|^{-2}$. Here, $T$ is the temperature, $k_\text{B}$ is Boltzmann's constant, and $\Delta\mu<0$ is the difference in chemical potential of the solid and the liquid. The proportionality depends on the interfacial tension and geometric factors due to the average shape of the critical nuclei.

\begin{figure}[t]
  \centering
  \includegraphics{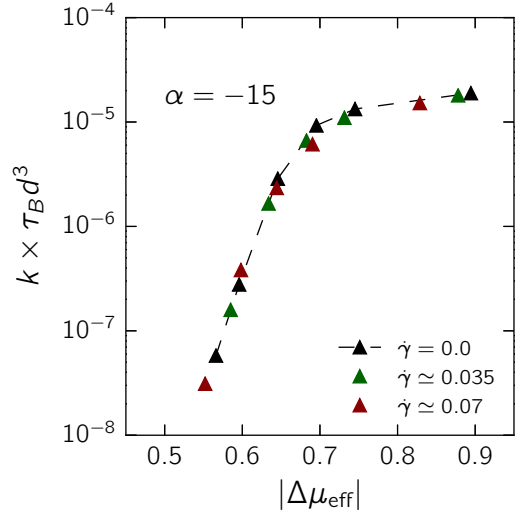}
  \caption{Effective chemical potential for packing fractions $\phi<\phi_0$. The crystallization rates $k$ for several strain rates collapse when plotted against an effective chemical potential. The dashed line is a guide to the eye.\label{fig:map}}
\end{figure}

Both contributions to the rate are affected when turning on the shear flow. The pre-factor is determined by the fluctuations and correlations between solid-like particles in the meta-stable liquid, the reduction of which is not strong enough to explain the observed exponential reduction of the crystallization rate. Alternatively, it has been proposed that the suppression can be accounted for through an effective free energy with an increased barrier~\cite{Blaak2004}. To test this idea, we convert packing fractions to chemical potential differences using a linear fit to the data of Ref.~\cite{dijkstra11}. Indeed, plotting the rates as a function of an effectively reduced chemical potential difference
\begin{equation}
  |\Delta\mu_\text{eff}| = |\Delta\mu|(1 + \alpha\sr^2) < |\Delta\mu|
\end{equation}
(while keeping the interfacial tension constant) with constant $\alpha\simeq-15$, all rates for different strain rates collapse onto a single curve as shown in Fig.~\ref{fig:map}. This demonstrates that nucleation under shear remains a rare, collective fluctuation to overcome a single barrier. Interestingly, Butler and Harrowell have argued for the non-equilibrium coexistence of a sheared liquid and the solid that the magnitude of the chemical potential difference is likely to increase~\cite{butl02,butl03}. This would imply that under shear there are two effective chemical potentials: one determines the nucleation of the new phase and one determines the final coexistence in the steady state. Both cannot be the same, which is in striking contrast to the quiescent liquid.

\subsection*{Diffusion-limited regime}

\begin{figure*}
  \centering
  \includegraphics[scale=1.0]{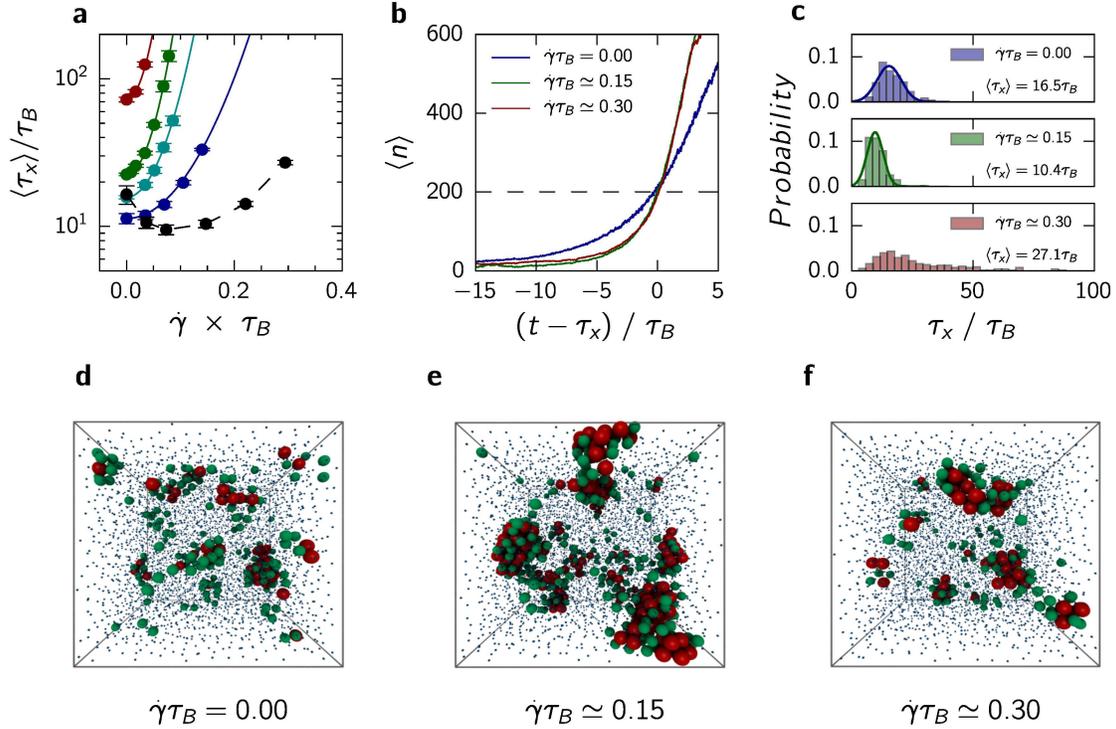}
  \caption{Shear-enhanced crystallization for $\phi\simeq0.587$. (a)~Mean crystallization time $\mean{\tx}$ \emph{vs.} the strain rate $\sr$ (black, other packing fractions are shown using the same colors as in Fig.~\ref{fig:rate}). The solid lines show the quadratic fits, the dashed line is a guide to the eye. (b)~Evolution of average cluster size $\mean{n}$ as a function of shifted time. (c)~Histograms of crystallization times for three strain rates and their respective means. Solid lines are Gaussian fits. (d-f)~Snapshots after $t=5\tbr$ starting from the same initial particle configuration for (d)~the quiescent liquid, (e)~at strain rate $\sr\tbr\simeq0.15$, and (f)~at strain rate $\sr\tbr\simeq0.3$. Red particles have been identified as solid-like, green particles as pre-structured (\emph{i.e.}, particles with high local order but less orientational ``bonds'' with their neighbors). For clarity, liquid-like particles are shown with a reduced size.\label{fig:enh}}
\end{figure*}

We finally study the regime $\phi\geqslant0.56$. In Fig.~\ref{fig:enh}a the mean crystallization time $\mean{\tx}$ is plotted for the largest packing fraction $\phi\simeq0.587$, which indeed shows an enhancement of crystallization for weak shear flow. For consistency, we maintain the same criteria for determining the crystallization time although now multiple solid clusters appear and grow. While at low supersaturation the distinction between the actual nucleation time (random waiting time to reach the critical cluster size) and crystallization time (random waiting time $\tx$ to reach $\ns$) is negligible, this is no longer the case. At this density, the solidification process is limited by the rate with which additional particles ``attach'' (there are no direct attractive forces but arranging into ordered structures locally increases the entropy). Shear flow increases the diffusion both parallel and perpendicular to the direction of the shear flow~\cite{land10}, and thus speeds up the attachment while the barrier to nucleate small solid clusters is still small. However, we observe that increasing $\sr$ the mean crystallization time reaches a minimum before it again increases for higher strain rates.

In Fig.~\ref{fig:enh}b the evolution of the largest cluster is shown for $\phi\simeq0.587$. To separate the effects of nucleation and growth, the single trajectories have been shifted by $\tx$ before averaging. In the quiescent liquid, the growth of the largest nucleus is gradually and slow before $\mean{n}$ becomes an approximately linear function of time for $\mean{n}>\ns$. Hence, the value $\ns=200$ is also a good estimate for the beginning of the linear growth regime. In the presence of shear flow, the initial increase of $\mean{n}$ is less gradual and the following growth is more rapid. Interestingly, for both non-zero strain rates the curves lie on top of each other while the crystallization rates differ by a factor of two. The reason is revealed in Fig.~\ref{fig:enh}c, which shows that the distribution of the single crystallization times $\tx$ is strongly affected by the shear flow. In the quiescent liquid the barrier to nucleation has basically vanished and we observe multiple nuclei (see snapshot Fig.~\ref{fig:enh}d). The mean nucleation time $\mean{\tx}\simeq16.5\tbr$ is small (compare $\mean{\tx}\simeq 750\tbr$ for $\phi\simeq0.539$) while the distribution is approximately Gaussian, which qualitatively agrees with a diffusive evolution under a constant force given by the chemical potential difference between liquid and solid. Going to a small strain rate $\sr\tbr\simeq0.15$, the distribution of nucleation times remains a Gaussian but becomes more narrow with a smaller mean. At this strain rate solid clusters thus appear more frequently and grow faster compared to the quiescent liquid (see Fig.~\ref{fig:enh}e). At higher strain rate $\sr\tbr\simeq0.3$ the distribution of $\tx$ becomes very broad with a pronounced tail, which implies a delayed induction of solid clusters and the qualitative change to activated behavior. Once these have formed, however, clusters grow faster than in the absence of shear flow (Fig.~\ref{fig:enh}f).

\section*{Conclusions}

The effect of weakly shearing a hard sphere liquid on crystallization strongly depends on the density: At low supersaturation the driving force of crystallization, the difference of the chemical potential between liquid and solid, is also small. Here the forces exerted by the shear flow might overcome the gain of entropy and lead to a larger probability for small solid clusters to lose particles (or even break up). Larger clusters are not affected, which, in combination, leads to a shift of the critical cluster size that can be interpreted as an effectively larger barrier to nucleation. No enhancement of crystallization occurs in this regime. This result might also shed some light onto the controversy regarding the discrepancy between experimental and simulation results at low supersaturations. As pointed out by Russo~\textit{et al.}~\cite{russ13}, sedimentation of not fully density-matched colloidal particles has an influence on the crystallization rate. However, the actual mechanism is not known and combined with the suppression of crystallization found here seems to rule out shear-induced effects.

At high supersaturation there is a combination of two effects: (i)~The diffusion of particles in the surrounding liquid is increased. This enhanced exploration of configuration space accelerates the growth of solid clusters. (ii)~The shear flow disrupts the subtle order in the liquid and thus suppresses the formation of small solid clusters. At larger strain rates this suppression is so strong that crystallization again becomes an activated process comparable to crystallization at lower supersaturation in the quiescent liquid. We expect that these physical mechanisms are valid beyond hard spheres also in the presence of short-ranged interactions.


\section*{Methods}

We study a model liquid composed of $N=5000$ monodisperse, nearly hard spheres in a periodic box with volume $V$ employing standard NVT molecular dynamics~\cite{Tild89}. Particles interact via the Weeks-Chandler-Andersen potential (WCA) $u(r)=4\epsilon [(\sigma/r)^{12}-(\sigma/r)^{6}+1/4 ]$ for $r<2^{1/6}\sigma$. The temperature is held constant via the stochastic Lowe-Andersen thermostat~\cite{Lowe06}, which is characterized by the bath collision frequency $\Gamma$. This thermostat has desirable properties, in particular it is Galilean invariant and conserves angular momentum. The WCA potential can be mapped onto hard spheres by means of an effective diameter $d$~\cite{dijkstra11}, which we compute from $\phi^\text{HS}_\text{f}=\frac{\pi}{6}\rho^\text{WCA}_\text{f}d^3$ with freezing packing fraction $\phi^\text{HS}_\text{f}\simeq0.492$ of hard spheres and freezing density $\rho^\text{WCA}_\text{f}\simeq0.712$ of the WCA liquid~\cite{dijkstra11}. The packing fraction is $\phi=\frac{\pi d^3}{6}\frac{N}{V}$. As time scale we employ the Brownian time $\tbr=d^2/D_0$, where $D_0$ is the bare diffusion coefficient in the infinitely dilute system. The system is driven through Lees-Edwards periodic boundary conditions leading to a linear flow profile $d\mean{v_x}/dy=\sr$.

Random configurations of dense hard spheres without preformed structures were created with the algorithm by Clarke and Wiley~\cite{Wiley87}, where the non-overlapping distance between particles was chosen to be equal to the effective diameter $d$. These configurations are thermalized in a short MD run without shear at high collision frequency $\Gamma\simeq100$ before production runs at given strain rate with $\Gamma\simeq10$.

To distinguish between liquid-like and solid-like particles, we follow Ref.~\citenum{dijkstra11}. We employ the local bond order parameter~\cite{Steinhardt1983}
\begin{equation}
  \label{eq:order}
  q_{l,m}(i) = \frac{1}{N_n(i)}\sum_{j=1}^{N_n(i)} Y_{l,m}(\theta_{i,j},\varphi_{i,j})
\end{equation}
for particle $i$, where $Y_{l,m}(\theta,\varphi)$ are spherical harmonics and $N_n$ is the number of neighbors within distance $r_{ij} < 1.5\sigma$. We construct a bond network through the scalar product
\begin{equation}
  \label{eq:scalar}
  d(i,j) = \frac{\sum_{m=-l}^{l} q_{l,m}(i) q_{l,m}^\ast(j) }{(\sum_{m=-l}^{l} |q_{l,m}(i)|)^{1/2} (\sum_{m=-l}^{l} |q_{l,m}(j)|)^{1/2}}
\end{equation}
using $l=6$ with $d(i,j)\geqslant0.7$ defining a bond. A particle is defined as ``solid-like'' (``pre-structured'') if the number of bonds is $\geqslant9$ ($\geqslant6$), and clusters are constructed from mutually bonded solid-like particles.



\section*{Acknowledgements}

We acknowledge financial support by DFG through collaborative research center TRR 146, and ZDV Mainz for computing time on the MOGON supercomputer.

\section*{Author contributions}

D.R. and T.S. designed the research and wrote the manuscript, D.R. wrote the code, performed simulations, and analyzed the data. All authors reviewed the manuscript.

\section*{Additional information}

The authors declare no competing financial interests.

\end{document}